\documentclass{PoS}
\usepackage{epsf}               %
\usepackage{latexsym}           %
\usepackage{amscd}              %
\usepackage{amsmath}            
\usepackage{graphicx,subfigure} %
\usepackage{comment}            %
\usepackage{cite}               
\input xy
\xyoption{all}

\renewcommand{\l}{\left}
\renewcommand{\r}{\right}
\renewcommand{\a}{\alpha}
\renewcommand{\b}{\beta}
\renewcommand{\c}{\gamma}

\newcommand{\beq}{\begin{equation}}
\newcommand{\eeq}{\end{equation}}
\newcommand{\bea}{\begin{eqnarray}}
\newcommand{\eea}{\end{eqnarray}}

\renewcommand{\o}{\overline}

\newcommand{\benn}{\begin{displaymath}}
\newcommand{\eenn}{\end{displaymath}}

\newcommand{\tw}{\textwidth}
\newcommand{\ig}{\includegraphics}
\def\slashchar#1{\ensuremath{                   %
   \setbox0=\hbox{${}#1{}$}                     
   \dimen0=\wd0                                 
   \setbox1=\hbox{/} \dimen1=\wd1               
   \ifdim\dimen0>\dimen1                        
   \rlap{\hbox to \dimen0{\hfil/\hfil}}         
   {}#1{}                                       
   \else                                        
   \rlap{\hbox to \dimen1{\hfil${}#1{}$\hfil}}  
   /                                            
   \fi}}                                        %
\def\simge{
 \mathrel{\rlap{\raise 0.511ex
  \hbox{$>$}}{\lower 0.511ex \hbox{$\sim$}}}}
\def\simle{
 \mathrel{\rlap{\raise 0.511ex
  \hbox{$<$}}{\lower 0.511ex \hbox{$\sim$}}}}
\title{Light baryon spectrum using improved interpolating operators}

\ShortTitle{Light baryon spectrum using improved interpolating operators}

\author{Lattice Hadron Physics Collaboration (LHPC): 
  S. Basak,$^a$ R. G. Edwards,$^b$ G. T. Fleming,$^c$ 
  J. Juge,$^d$ A. Lichtl,$^d$ C. Morningstar$^d$ 
  D. G. Richards,$^b$ \speaker{I. Sato},$^e$
  S. J. Wallace,$^f$\\
\llap{$^a$}Department of Physics, N.N.D. College, Calcutta 700092, India\\
\llap{$^b$}Thomas Jefferson National Accelerator Facility, Newport News, VA 23606, USA\\
\llap{$^c$}Yale University, New Haven, CT 06520, USA\\
\llap{$^d$}Department of Physics, Carnegie Mellon University, Pittsburgh, PA 15213, USA\\
\llap{$^e$}Lawrence Berkeley Laboratory, Berkeley, CA 94720, USA\\
\llap{$^f$}Department of Physics, University of Maryland, College Park, MD 20742, USA}

\abstract{Energies for excited light baryons are computed in
quenched QCD with a pion mass of 490\,MeV.
Operators used in the simulations include local operators and
the simplest nonlocal operators that have nontrivial orbital structures.
All operators are designed with the use of Clebsch-Gordan
coefficients of the octahedral group so that they transform
irreducibly under the group rotations.
Matrices of correlation functions are computed for each irreducible
representation, and then the variational method is applied to separate
mass eigenstates.
We obtained 17 states for isospin 1/2 and 11 states for isospin 3/2
in various spin-parity channels including $J^P=5/2^\pm$.
The pattern of the lowest-lying energies from each irrep is discussed.
We use anisotropic lattices of volume $24^3\times 64$ 
with temporal lattice spacing $a_t^{-1}=6.05$\,GeV with renormalized 
anisotropy $\xi=3.0$.}

\FullConference{XXIVth International Symposium on Lattice Field Theory\\
                July 23-28, 2006\\
                Tucson, Arizona, USA}

\begin{document}

\section{Introduction}

The theoretical determination of the baryon spectrum is an important
step for lattice QCD.
Lattice QCD results have provided reasonable agreement with the 
physical ground state masses of different baryons (for a review, 
see~\cite{Ishikawa:2004nm}).
For excited states of baryons, several groups have reported 
results~\cite{Zhou:2006xe,Burch:2006cc,Sasaki:2005uq,
Sasaki:2005ug,Basak:2004hr,Zanotti:2003fx}, but further improvements
are needed for more precise determinations of the masses and spins
for excited particles.
In this report, we show preliminary results for nucleon and delta baryon masses
including $J^P=5/2^\pm$ using improved lattice baryon fields
with quenched anisotropic lattices.

Spins of states on a cubic lattice are deduced from the 
level orderings of energies computed in different irreducible
representations (irreps) of the octahedral group.
Source and sink operators have to be designed so as to transform as 
a definite irrep of the octahedral group in order to identify 
a total angular momentum of a continuum state.
There are six double-valued irreps of the octahedral group: 
$G_{1g}, H_g, G_{2g}, G_{1u}, H_u$ and $G_{2u}$, in which
``g'' subscript denotes positive parity ({\it gerade}) and
``u'' subscript denotes negative parity ({\it ungerade}).
Subduction pattern of total angular momentum to octahedral irreps
is summarized in Table~\ref{table:subduction}. 

In general, a lattice operator with a definite irrep couples
to multiple states corresponding to different total angular momenta according
to the subduction pattern.
To separate the individual energy eigenstates included in an
irrep, we employ the variational technique, which involves diagonalization
of a matrix of correlation functions.
Substantial numbers of basis operators are needed in each channel
in order to provide sufficiently large degrees of freedom 
for the determination of the spectrum of excited states.
Inclusion of multi-hadron operators in the construction of matrices of correlation 
functions is needed to separate the resonances from the hadron scattering states.
In this work, we only used three-quark operators. Thus the separation
of discrete scattering states and single-hadron states in the excited 
spectrum is beyond scope of this work, and we refer to our lattice results
as ``energies'' rather than as ``masses''.
\begin{table}[h]
\caption{The number of occurrences of double-valued irrep 
$\Lambda=\{G_1,H,G_2\}$ of the octahedral group for
different values of continuum $J$ (up to $11/2$).}
\label{table:subduction}
\begin{center}
\begin{tabular}{|cllllllll|}
\hline
$\Lambda$&  &J=& ${1\over 2}$&${3\over 2}$& ${5\over 2}$
               & ${7\over 2}$& ${9\over 2}$& ${11 \over 2}$ \\
$G_{1}$  &  &  & 1 & 0 & 0 & 1 & 1 & 1  \\
$H $     &  &  & 0 & 1 & 1 & 1 & 2 & 2  \\
$G_{2}$  &  &  & 0 & 0 & 1 & 1 & 0 & 1  \\
\hline
\end{tabular}
\end{center}
\end{table}

\section{Improved baryon operators}

We construct operators that transform according to each irrep of the
octahedral group by employing the analytical method based on appropriate
Clebsch-Gordan coefficients for the group~\cite{Basak:2005ir}.
An alternative method of constructing vast numbers of operators in
an automated procedure is presented in Ref.~\cite{Basak:2005aq}.
We used complete sets of quasi-local and one-link three-quark 
operators in this work.

%
Quasi-local baryon operators consist of smeared quark fields 
$\widetilde{q}({\bf x}, t)$ that
are located at the same space point at which the color singlet
contraction is taken,
\begin{equation}
\o\Psi^{(IS)}_{\Lambda,\lambda,m}({\bf x},t)=
\epsilon_{ijk} f^{(IS)}_{abc} c^{(\Lambda,\lambda,m)}_{\alpha \beta \gamma}
\o{\widetilde{q}}^{i,a}_\alpha({\bf x},t) 
\o{\widetilde{q}}^{j,b}_\beta({\bf x},t) 
\o{\widetilde{q}}^{k,c}_\gamma({\bf x},t),
\end{equation}
where $i,j,k$ are color indices, $a,b,c$ are flavor indices, and
$\alpha,\beta,\gamma$ are Dirac indices.
Octahedral group irreps are denoted by 
$\Lambda=\{ G_{1g},H_{g},G_{2g},G_{1u},H_{u},G_{2u}\}$.
The corresponding dimensions $d_\Lambda$ are 2, 4, 2, 2, 4, 2.
The rows of an irrep are distinguished by row labels $\lambda=1,\cdots,d_\Lambda$.
When there are more than one operators in a single 
$(\Lambda,\lambda)$ channel, we use the additional index $m$ to label
the different operators. These labels are the indices of matrices of
correlation functions when source and sink operators from a given irrep
are used.
Isospin $I$ and strangeness $S$ are selected by coefficients $f^{(IS)}_{abc}$.
We use the mass for up and down quarks, and do not consider 
strange quarks in this work.
The explicit Clebsch-Gordan coefficients 
$c^{(\Lambda,\lambda,m)}_{\alpha\beta\gamma}$ are given in Ref.~\cite{Basak:2005ir}.
A simple extension of quasi-local operators is 
to displace one of the quark fields along a spatial axis and
to include a straight gauge-link so as to keep the gauge covariance.
The displacement operator $\hat{d}_l$ acted on three-quark operator
is defined as follows,
\begin{align}
\hat{d}_l \, &\o\Psi^{(IS)}_{\Lambda,\lambda,k}({\bf x},t) 
= \epsilon_{ijk} f^{(IS)}_{abc}
c^{(\Lambda,\lambda,k)}_{\a\b\c} 
\o{\tilde{q}}_\a^{i, a}({\bf x},t)
\o{\tilde{q}}_\b^{j, b}({\bf x},t)
\o{\tilde{q}}_\c^{k',c}({\bf x}+a_s\hat{l},t) U^{\dag k'k}_{l}({\bf x},t),
\end{align}
where $\hat{l}$ takes six spatial directions, $\pm \hat{x}, \pm \hat{y}, \pm \hat{z}$.
Assuming that lattice gauge links recover the cubic symmetry 
with large statistics,
displacement operators transform amongst themselves under lattice rotations.
By having the following similarity transformed bases,
\begin{equation}
\begin{pmatrix}
  \hat{A}_1 \o\psi \\
  \hat{D}_+ \o\psi \\
  \hat{D}_- \o\psi \\
  \hat{D}_0 \o\psi \\
  \hat{E}_0 \o\psi \\
  \hat{E}_2 \o\psi 
\end{pmatrix}
\equiv 
\begin{pmatrix}
  {1\over \sqrt{6}}( \hat{d}_x\o\psi + \hat{d}_y\o\psi + 
  \hat{d}_z\o\psi + \hat{d}_{-x}\o\psi + 
  \hat{d}_{-y}\o\psi + \hat{d}_{-z}\o\psi) \\
      {i \over 2}[ (\hat{d}_x \o\psi - \hat{d}_{-x}\o\psi)
	+ i( \hat{d}_y \o\psi - \hat{d}_{-y}\o\psi)] \\
      -{i\over 2}[(\hat{d}_x\o\psi - \hat{d}_{-x}\o\psi)
	- i(\hat{d}_y\o\psi - \hat{d}_{-y}\o\psi)] \\
      -{i\over \sqrt{2}}(\hat{d}_z\o\psi - \hat{d}_{-z}\o\psi)\\
    \! {1\over \sqrt{12}}[2(\hat{d}_z\o\psi + \hat{d}_{-z}\o\psi)
      -(\hat{d}_x\o\psi + \hat{d}_{-x}\o\psi) - 
      (\hat{d}_y\o\psi + \hat{d}_{-y}\o\psi)] \! \\
    {1\over 2}[(\hat{d}_x\o\psi + \hat{d}_{-x}\o\psi) - 
      (\hat{d}_y\o\psi + \hat{d}_{-y}\o\psi)]
\end{pmatrix}
\end{equation}
the displacements $\hat{A}_1, \hat{D}_{\pm,0}$, and $\hat{E}_{0,2}$ 
transform according to $A_1$, $T_1$, and $E$ irreps of the octahedral group, 
respectively.
Here $A_1$ is the cubically symmetric, one-dimensional irrep,
$T_1$ is the three-dimensional irrep that corresponds to $\ell=1$ and
$E$ is the two-dimensional irrep that corresponds to $\ell=2$ for the
lowest orbital angular momentum $\ell$.
These new bases are chosen so that their transformations resemble 
those of the spherical harmonics $Y_{\ell m}$, {\it i.e.},
$\hat{A}_1 \sim Y_{00}$, 
$\hat{D}_{+,0,-} \sim Y_{11}, Y_{10}, Y_{1-1}$, and
$\hat{E}_{0,2} \sim Y_{20}, (Y_{22}+Y_{2-2})$.
The $A_1$, $T_1$, and $E$ one-link operators are now defined as
\begin{align}
\o B^{(IS)}_{\Lambda,\lambda,m}({\bf x},t)= \l\{
\begin{array}{l}
\hat{A}_1 \o\Psi^{(IS)}_{\Lambda,\lambda,m}({\bf x},t) \\
\sum_{r,\lambda'} C
\renewcommand{\arraystretch}{0.6}
\begin{pmatrix}\Lambda & T_1 & \Lambda' \\ \lambda & r & \lambda'\end{pmatrix}
\hat{D}_r \o\Psi^{(IS)}_{\Lambda',\lambda',m'}({\bf x},t) \\
\sum_{r,\lambda'} C
\renewcommand{\arraystretch}{0.6}
\begin{pmatrix}\Lambda & E & \Lambda' \\ \lambda & r & \lambda'\end{pmatrix}
\hat{E}_r \o\Psi^{(IS)}_{\Lambda',\lambda',m'}({\bf x},t)
\end{array}
\r. 
~~~\equiv c^{(\Lambda,\lambda,m)}_{\alpha\beta\gamma, i} \hat{d}_i 
\o{B}_{\alpha\beta\gamma},
\end{align}
where 
$C
\renewcommand{\arraystretch}{0.6}
\begin{pmatrix}
\Lambda & \Lambda' & \Lambda'' \\ 
\lambda & \lambda' & \lambda''
\end{pmatrix}$
are the Clebsch-Gordan coefficients that takes the direct product of 
irreps $\Lambda'$ and $\Lambda''$ to yield overall irrep 
$\Lambda$~\cite{Basak:2005ir}.

We then use improved operators to construct matrices of correlation functions 
\begin{align}
C_{mm'}^{(\Lambda\lambda)}(t) \delta_{\Lambda\Lambda'} \delta_{\lambda\lambda'}=
c^{(\Lambda\lambda m)*}_{\a\b\c,i} c^{(\Lambda'\lambda' m')}_{\mu\nu\rho,i'}
\sum_{\bf x} 
\langle 0 \vert B_{\a\b\c}({\bf x},t) \hat{d}^\dag_i
\hat{d}_{i'} \o B_{\a'\!\b'\!\c'\!}({\bf 0},0) \vert 0 \rangle
\gamma^4_{\a' \mu}\gamma^4_{\b' \nu}\gamma^4_{\c' \rho},
\label{eq:matrix_correlation_function}
\end{align}
where a displacement operator $\hat{d}_i$ applied to a baryon field
denotes a one-link baryon operator for $i=\pm x, \pm y, \pm z$
and a quasi-local baryon for $i=0$.
Lattice operators belonging to different irreps or to different
rows of one irrep are orthogonal because of the octahedral symmetry of the lattice.
Different embeddings of a given irrep and row provide 
operators appropriate to forming a matrix of correlation functions,
to which the variational method is applied
in order to determine the spectrum of states.
Three Dirac $\gamma_4$ matrices are included  
in Eq.~(\ref{eq:matrix_correlation_function}) in order to produce a
Hermitian matrix of correlation functions.

\section{Computational methods}

\subsection{Variational method}

Analyses of excited state energies are based upon the matrices of
correlation functions of
Eq.~(\ref{eq:matrix_correlation_function}).
In order to extract the spectrum of energies from the matrix of
correlation functions, we first average over rows because they
provide equivalent results owing to octahedral symmetry.  We then solve the following
generalized eigenvalue equation,
\begin{equation}
\sum_{k'} \widetilde{C}^{(\Lambda)}_{kk'}(t) v^{(n)}_{k'}
(t,t_0) = \alpha^{(n)}(t,t_0) \sum_{k'}
\widetilde{C}^{(\Lambda)}_{kk'}(t_0) v^{(n)}_{k'}(t,t_0),
\label{eq:new_generalized_eigenvalue_equation}
\end{equation}
where superscript $n$ labels the eigenstates.  The symbol $\widetilde{C}^{(\Lambda)}$
indicates that the appropriate average over rows has been performed. 
The reference time $t_0$ in
Eq.~(\ref{eq:new_generalized_eigenvalue_equation}) is taken near the
source time $t=0$ in order to have significant contributions from
excited states. The generalized eigenvalues $\alpha^{(n)}(t,t_0)$
are related to the energy $E_n$ by~\cite{Luscher:1990ck}
\begin{equation}
\alpha^{(n)}(t,t_0) \simeq e^{-E_n(t-t_0)}[1+O(e^{-|\delta E|t})], 
\label{eq:generalized_eigenvalue}
\end{equation}
where $\delta E$ is the difference between $E_n$ and the next
closest energy.
 We have determined effective energies $E_n$ by fitting the
generalized eigenvalues to the leading term of Eq.~(\ref{eq:generalized_eigenvalue}).  

\subsection{Lattice action}

We employed anisotropic lattices with volume $24^3\times 64$, where
the temporal lattice spacing $a_t$ is three times finer than the 
spatial lattice spacing $a_s$.  
Gauge configurations are generated using the quenched, anisotropic,
unimproved Wilson gauge 
action~\cite{Klassen:1998ua}
with $\beta = 6.1$ and the pion mass is 490 MeV.  
A total of 167 configurations is used, but statistical ensembles are 
effectively doubled by use of CPT symmetry.
The scale is $a^{-1}_t = 6.05$\,Gev in these lattice,
provided by the analysis of string tension.

\section{Results and discussions}

\subsection{Effective energies}

\begin{figure}[h]
\centering
\ig[width=\tw]{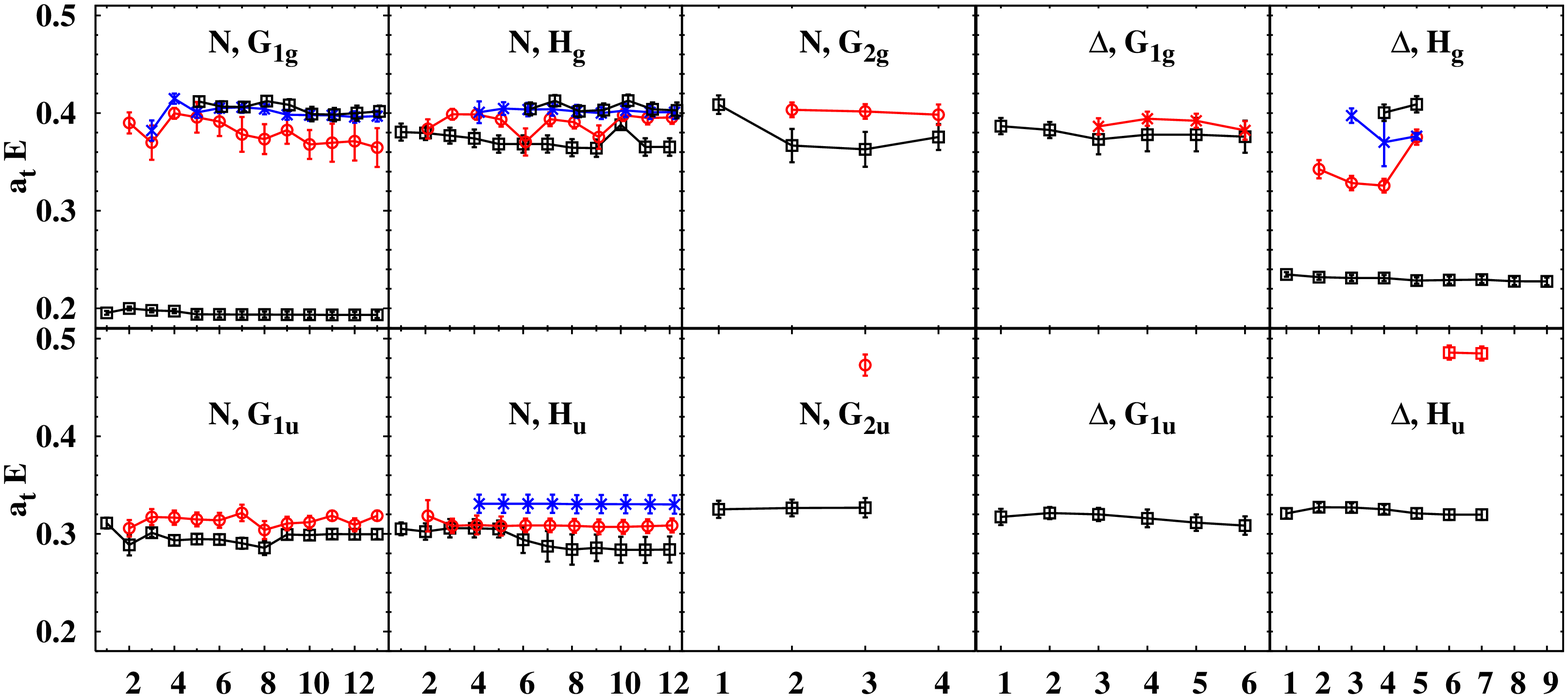} 
\caption{Effective masses in lattice units
vs.\ dimension of matrices of correlation functions.
Upper panels show the positive-parity channels and 
lower panels show the negative-parities channels.
no matrices of correlation functions are 
obtained for the $\Delta, G_{2g/u}$ irreps.}
\label{fig:stability}
\vspace{3mm}
\centering 
\ig[width=0.51\tw]{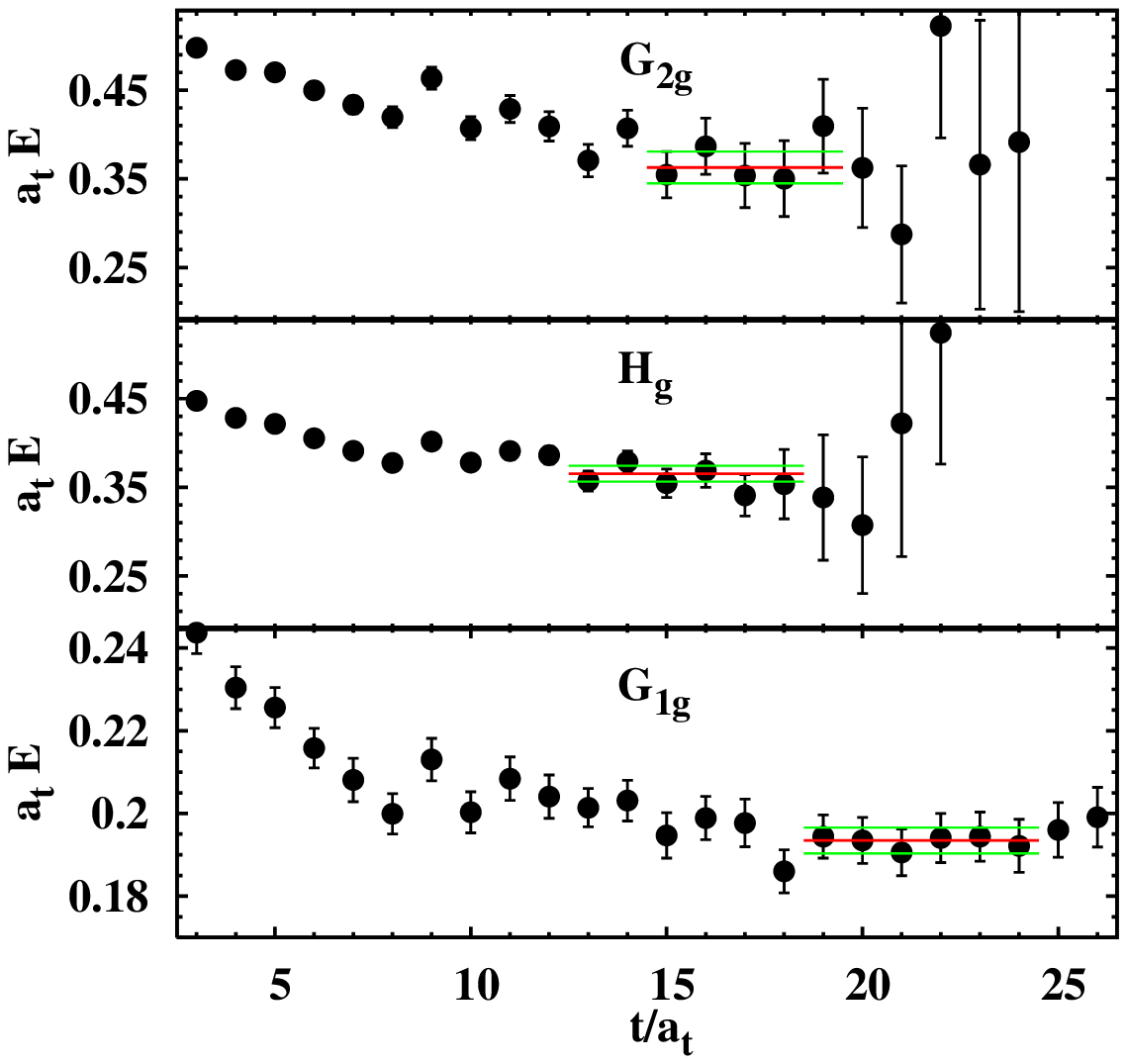}
\caption{Effective energies for the lowest states in $G_{1g}, H_{g}$ and 
$G_{2g}$ irreps with isospin 1/2.
Time ranges used for and energies obtained from fits are shown by horizontal lines. } 
\label{fig:EffMass_N24}
\end{figure}

We have extracted energies for isospin $1/2$ and $3/2$
channels by
diagonalizing matrices of correlation functions formed from
three-quark operators that share the same octahedral 
symmetry. 
We diagonalize matrices of different
dimensions by adding or subtracting trial operators to
find optimal sets of operators.
Figure~\ref{fig:stability} shows low-lying effective energies 
as a function of matrix dimensions.  Energies are determined
from single-exponential fits to the generalized eigenvalues.  
We found that low-lying energy states are fairly stable with respect to 
the number of operators used, as long as a few important operators 
are included.
It is expected that use of a larger number of operators should improves 
the low-lying spectrum because contaminations from higher-lying 
states are reduced.
This expectation appears to hold except for a few excited states
that are sensitive to the types of operators used.
The lowest-lying effective energies in the $G_{1g},H_g,G_{2g}$
channels with isospin 1/2 are plotted in Fig.~\ref{fig:EffMass_N24}.

\subsection{Pattern of low-lying energies}

\begin{figure}[b]
\centering 
\ig[width=1.0\tw]{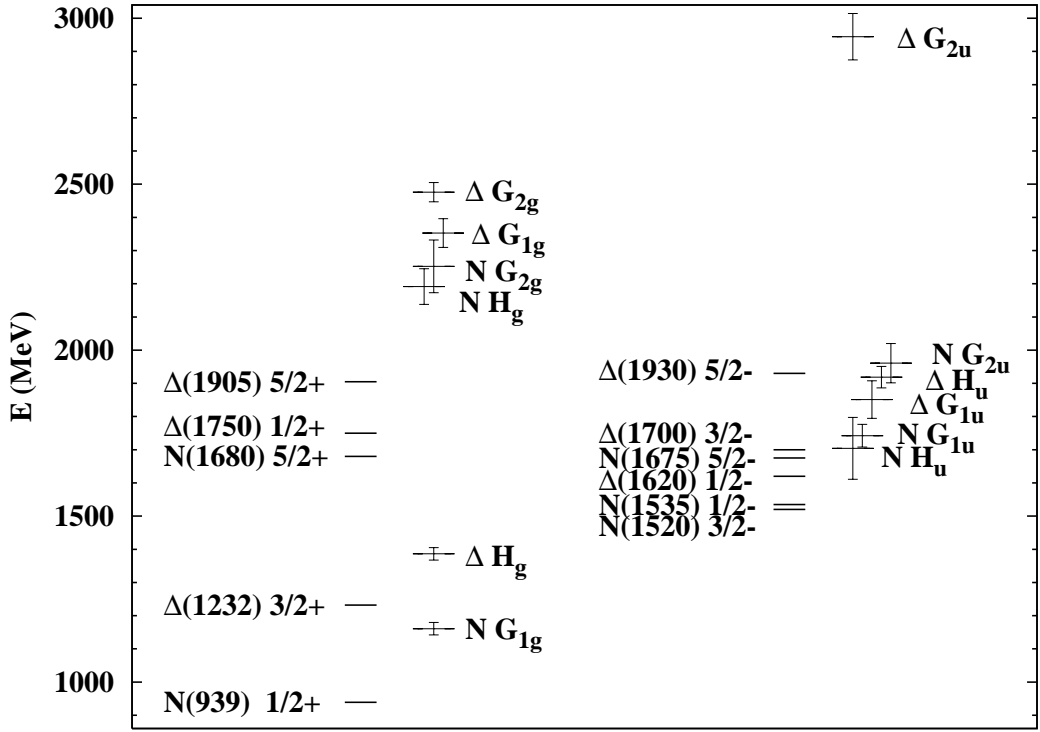} 
\caption{The energies obtained for each symmetry channel 
of isospin $1/2$ and $3/2$ baryons are shown.
The energies are converted in MeV using the scale $a_t^{-1}=6.05$\,GeV.}
\label{fig:N24energies}
\end{figure}
Figure~\ref{fig:N24energies} show the lowest energies from each irrep
in isospin 1/2 and 3/2.  Energies are converted to physical units.
They are determined based the diagonalization of matrices of correlation 
functions consisting of the best sets of operators.
This graph shows the pattern of level orderings for the
different irreps, but it is not to compare the results with experimental values
because our pion mass is too heavy and there may be significant 
discretization errors.
Nevertheless, it is interesting to compare the
{\it pattern} of level orderings with empirical data.

In the positive-parity channels, besides the nucleon ground state and 
the delta ground state, isospin 1/2, $G_{2g}$ and $H_g$ states are nearly
degenerate within errors.
If this degeneracy holds in the continuum limit and no degenerate partner 
exists in $G_{1g}$, these two eigenstates correspond to the subduction of
$J^P=5/2^+$.
In our simulations we have no evidence for a $3/2^+$ state that has lower
energy than $5/2^+$ state.
Indeed in nature $N(5/2^+, 1675)$ is lighter than $N(3/2^+, 1705)$.
Our results agree with this pattern, though our 
simulation conditions are far from physical.
Previous lattice calculations that used $H_g(H_u)$ operators, such as
Rarita-Schwinger projected operators, assumed that the lightest
state corresponds to spin $3/2^+(3/2^-)$.
Spin identification of the lowest $H$ state is uncertain unless $G_2$ irreps
are included, i.e., the state could have spin 5/2 or higher.
The $G_2$ baryon operators cannot be constructed by local operators,
therefore displaced operators are needed for the lattice simulations of
excited baryon spectrum.

In the negative-parity channels, the level orderings from the lattice 
simulations go as follows:
the $N^*, H_u$ corresponding to $J^P=3/2^-$ (no degenerate partner with $N^*, G_{2u}$) 
is the lowest, 
the $N^*, G_{1u}$ corresponding to $1/2^-$ is slightly above, 
the $\Delta^*, G_{1u}$ corresponding to $1/2^-$ is next, 
the $\Delta^*, H_u$ corresponding to $3/2^-$ 
(no degenerate partner with $\Delta^*, G_{2u}$)
is slightly above, 
the $N^*, G_{2u}$ corresponding to $5/2^-$ is next and finally 
the $\Delta^*, G_{2u}$ corresponding to $5/2^-$ is the highest.
These spin identifications are only tentative.
In our simulations, we cannot find more that one operator in isospin 3/2,
$G_{2g/u}$ channel, thus there is uncertainty in these channels.
The pattern of level orderings in the negative-parity channels is 
consistent with the experimental
data, except for the ordering of the $\Delta^*, H_u$ and the $N^*, G_{2u}$ states.

It is intriguing to note that comparison of energies between different parities 
gives very different conclusions.
For instance, the (lowest) $\Delta^*, 3/2^-$ state should appear between
the $N^*, 5/2^+$ and the $\Delta^*, 1/2^+$, but our calculation
does not agree with this pattern.
We do not find a positive-parity excited state that has lighter energy than
the lowest negative-parity state in isospin 1/2.
Further studies of lattice simulations are needed to understand 
realistic mass gaps between parities.

For more details about the quenched baryon spectrum based
on extended operators, readers are referred to Ref.~\cite{Lichtl-Thesis}.

This work was supported by the U.S. National Science Foundation under
Award PHY-0354982 and by the U.S. Department of Energy
under contracts DE-AC05-06OR23177 and DE-FG02-93ER-40762.

\end{document}